
\magnification \magstep1
\raggedbottom
\openup 4\jot
\voffset6truemm
\headline={\ifnum\pageno=1\hfill\else
\hfill{\it Singularity Theory in Classical Cosmology}
\hfill \fi}
\rightline {January 1992, DSF-92/6}
\centerline {\bf SINGULARITY THEORY IN CLASSICAL COSMOLOGY}
\vskip 1cm
\leftline {GIAMPIERO ESPOSITO}
\leftline {\it Istituto Nazionale di Fisica Nucleare, Sezione
di Napoli, Gruppo IV,}
\leftline {\it Mostra d'Oltremare, Padiglione 20, 80125 Napoli}
\leftline {\it Dipartimento di Scienze Fisiche,}
\leftline {\it Mostra d'Oltremare, Padiglione 19, 80125 Napoli}
\vskip 1cm
\noindent
{\bf Summary. -} This paper compares recent approaches
appearing in the literature on the singularity problem
for space-times with nonvanishing torsion.
formalism.
of gravitation.
\vskip 10cm
In the last few years, a new approach has been proposed by
the author to singularity theory in classical cosmology
with nonvanishing torsion, based on the definition of
geodesics as curves whose tangent vector moves by parallel
transport [1,2]. With the notation described in Refs.
[1,2], our main result can be stated as follows.
\vskip 0.3cm
\noindent
{\bf Theorem.} The $U_{4}$ space-time of the ECSK theory
cannot be timelike geodesically complete if :

(1) $\Bigr[R(U,U)-2{\widetilde \omega}^{2}\Bigr] \geq 0$
for any nonspacelike vector $U$;

(2) there exists a compact spacelike three-surface $S$
without edge;

(3) the trace of the extrinsic curvature tensor of $S$
is either everywhere positive or everywhere negative
[this tensor also plays a role in the theory of maximal
timelike geodesics and partial Cauchy surfaces].

However, the generalization of singularity theorems to
$U_{4}$ space-times has also been addressed in Ref. [3].
The differences between our work and these papers are :

(1) The author of Ref. [3] does not define geodesics as
autoparallel curves;

(2) After a review of the approach in Ref. [4], he still
makes a splitting so as to express
$R(U,U) \equiv R_{ab}U^{a}U^{b}$ as the part formally
identical to general relativity plus other contributions
involving torsion;

(3) He derives the Hawking-Penrose timelike convergence
condition in $U_{4}$ space-time for a shear-free and
convergence-free congruence, and he obtains
$R(U,U) \geq 0$, since he separately requires
(using the notation in Eqs. (5.2-3) of Ref. [1])
$$
{1\over 4} {\widetilde S}_{ab}{\widetilde S}^{ab}
\geq \left[{\theta^{2} \over 3}+{d \theta \over ds}
\right]
\; \; \; \; .
\eqno (1)
$$
However, in the generalized Raychaudhuri equation,
$R(U,U)$ and $2{\widetilde \omega}^{2}$ occur with
opposite signs, so that one has to require in general
condition (1) of our theorem, if
${d\theta \over ds}$ has to remain less than or equal
to $-{\theta^{2}\over 3}$, as we explained in Refs.
[1,2]. Thus, in our work (see also Ref. [5]), torsion
explicitly appears in writing down
${\widetilde \omega}^{2}$ appearing in condition (1)
of the theorem stated above, whereas Eq. (5.8) of
Ref. [1] is written as in Einstein's theory, and the
singularity theorem is proved under only three
assumptions, as in general relativity. By contrast, in
the work appearing in Ref. [3], one requires two
separate conditions : $R(U,U) \geq 0$ and Eq. (1)
of this note, instead of condition (1) of our theorem.

(4) The author of Ref. [3] does not consider the full
extrinsic curvature tensor and conditions for maximal
timelike geodesics, and he does not avoid the introduction
of a modified energy-momentum tensor;

(5) The author of Ref. [3] studies the generalization of the
Hawking-Penrose singularity theorem.

However, appendix A in the first paper of Ref. [3] deals
with Hawking's singularity theorem without causality
assumptions in general relativity, and in the concluding
sect. {\bf 4} of that paper it is emphasized that one
could investigate singularities in space-times with torsion
by looking at the completeness of autoparallels (which are
there called nongeodesic curves). It is thus possible that
the derivation of our theorem, first published in Ref. [1],
will be studied independently by other authors in the near
future.

Another interesting study of the inclusion of spin in
the Raychaudhuri equation can be found in Ref. [6], where
this equation is applied to the behaviour of an
irrotational, unaccelerated fluid, and the development
of singularities in the expansion is studied for
constant spin densities. The fundamental difference between
our work and their work is the following. In Eq. (19)
of Ref. [6], a Raychaudhuri equation is written for
space-times with torsion where all covariant derivatives
only contain Christoffel symbols. The spin connection
has been separated from the covariant derivatives and
explicitly included. However, as we already emphasized
in Refs. [1,2], this split is not in agreement with the
Hamiltonian treatment of theories with torsion. Hence
we can repeat the remarks made in comparing our work
with Ref. [3].

We should emphasize this paper is not aiming
at criticizing Refs. [3,6], and we only
hope to stimulate further research on singularity theory
in classical cosmology with torsion. We are grateful to
Dr. P. Scudellaro for bringing Ref. [3] to our
attention.
\vskip 0.3cm
\leftline {\it REFERENCES}
\item {[1]}
G. ESPOSITO: {\it Nuovo Cimento B}, {\bf 105}, 75 (1990);
G. ESPOSITO: {\it Nuovo Cimento B}, {\bf 106}, 1315 (1991).
\item {[2]}
G. ESPOSITO: {\it Fortschritte der Physik}, {\bf 40}, 1 (1992).
\item {[3]}
L. C. GARCIA DE ANDRADE: {\it Found. Phys.},
{\bf 20}, 403 (1990); L. C. GARCIA DE ANDRADE:
{\it Intern. J. Theor. Phys.}, {\bf 29}, 997 (1990).
\item {[4]}
F. W. HEHL, P. VON DER HEYDE and G. D. KERLICK:
{\it Phys. Rev.}, {\bf D 10}, 1066 (1974).
\item {[5]}
J. M. STEWART and P. H\`AJICEK: {\it Nat. Phys.
Sci.}, {\bf 244}, 96 (1973).
\item {[6]}
A. J. FENNELLY, J. P. KRISCH, J. R. RAY and L. L.
SMALLEY: {\it J. Math. Phys.}, {\bf 32}, 485 (1991).
\bye